\documentclass[journal=jacsat,manuscript=article]{achemso}

\usepackage{chemformula} 
\usepackage[T1]{fontenc} 

\newif\ifComment
\Commentfalse
\graphicspath{{./Figs/}}

\newcommand{\bs}[1]{\boldsymbol{#1}}

 \SectionNumbersOn
\author{Emadeldeen Hassan}
\email{emadeldeen.hassan@umu.se}
\affiliation[LUH]{Hannover Centre for Optical Technologies, Institute for Transport and Automation Technology (Faculty of Mechanical Engineering), and Cluster of Excellence PhoenixD, Leibniz University Hannover, 30167 Hannover, Germany}
\alsoaffiliation[MNF]
{Department of Electronics and Electrical Communications, \\ Menoufia University, Menouf 32952, Egypt}
\alsoaffiliation[UMU]
{Department of Applied Physics and Electronics, \\Ume\aa \mbox{} University, SE-901~87~Ume{\aa}, Sweden}
\author{ Antonio Cal\`a Lesina}
\affiliation[LUH]{Hannover Centre for Optical Technologies, Institute for Transport and Automation Technology (Faculty of Mechanical Engineering), and Cluster of Excellence PhoenixD, Leibniz University Hannover, 30167 Hannover, Germany}
\email{antonio.calalesina@hot.uni-hannover.de}

\title[An \textsf{achemso} demo]
  {Topology optimization of dispersive plasmonic nanostructures in the time-domain}

\abbreviations{IR,NMR,UV}

\begin{document}

%
%
%
%
%

\begin{abstract}
Topology optimization techniques have been applied in integrated optics and nanophotonics for the inverse design of devices with shapes that cannot be conceived by human intuition. 
At optical frequencies, these techniques have only been utilized to optimize nondispersive materials using frequency-domain methods.
However, a time-domain formulation is more efficient to optimize materials with dispersion. 
We introduce such a formulation for the Drude model, which is widely used to simulate the dispersive properties of metals, conductive oxides, and conductive polymers. 
Our topology optimization algorithm is based on the finite-difference time-domain (FDTD) method, and we introduce a time-domain sensitivity analysis that enables the evaluation of the gradient information by using one additional FDTD simulation. 
The existence of dielectric and metallic structures in the design space produces plasmonic field enhancement that causes convergence issues. 
We employ an artificial damping approach during the optimization iterations that, by reducing the plasmonic effects, solves the convergence problem. 
We present several design examples of 2D and 3D plasmonic nanoantennas with optimized field localization and enhancement in frequency bands of choice.
Our method has the potential to speed up the design of \emph{wideband} optical nanostructures made of dispersive materials for applications in nanoplasmonics, integrated optics, ultrafast photonics, and nonlinear optics.
\end{abstract}


\section{Introduction}
The last decade has witnessed an exponential increase of research in nanophotonics.
Plasmonic and dielectric nanostructured materials, such as metasurfaces and metamaterials, have been developed to engineer the properties of light beyond what is allowed by bulk optical devices\cite{Yu2011a,Kamali2018}, thus leading to revolutionary solutions for beam structuring\cite{karimi14generating}, colouring\cite{guay17Laser}, biosensing\cite{Tseng2021}, nanomedicine\cite{Baffou2020}, tunable beam steering\cite{Lesina21Tunable}, and nonlinear generation\cite{Lee2014,Lesina17Vectorial}, to name a few. 
Advances in nanofabrication technologies have enabled such technologies by allowing unprecedented design complexity at the nanoscale\cite{Rashid21Helium,Tseng2021}. The opportunities offered by theoretical research and nanofabrication facilities raise the need for new methods to efficiently design and optimize such nanophotonic systems\cite{molesky18inverse,Campbell19Review}.

Advances in computing capabilities and numerical methods, such as the finite-difference time-domain (FDTD)\cite{Taflove05} and finite-element methods (FEM)\cite{jin14FEM}, empowered the design of nanophotonic devices while shortening their design cycle.
Conventionally, a design cycle starts from a given layout.
The layout is then parameterized and various techniques are employed to explore the parameters' space to find satisfying solutions.
The use of large parameter spaces offers opportunities to find new designs with improved performance or designs that can satisfy multi-objectives.
However, exploring such large parameter spaces raises computational challenges.
Techniques such as parameters sweep or stochastic optimization methods, e.g., genetic algorithms, are computationally intractable to explore large design spaces and are only suitable to handle problems with few design parameters\cite{Sigmund11On}. Deep learning algorithms require large data sets for training, and their use in the inverse design of nanophotonics is still in its infancy\cite{baxter19plasmonic,Chen20Physics,ma21deep}.

Topology optimization (TopOpt) is a robust inverse design approach\cite{Bendsoe2004,Deaton2014survey}.
It was initially introduced to optimize mechanical structures\cite{BENDSOE88}, then it has been successfully extended to various engineering disciplines including acoustics\cite{WADBRO06Topoloy}, fluids\cite{GersborgHansen2005}, and electromagnetics\cite{Nomura07,Hassan14Topology,Aage017Topology,wang17Antenna,Hassan20Multilayer}.
Typically, TopOpt problems are solved using gradient-based optimization methods, where the gradient of the objective function is computed using efficient methods such as the adjoint-field method\cite{zhang_wideband_14,Hassan15Time}.
Optimization problems that include millions or even billions of design variables have enabled novel conceptual designs\cite{Aage17Giga}.
In electromagnetics, TopOpt was used to optimize non-dispersive dielectric devices in the microwave and optical regimes\cite{Jensen2010,ELESIN14Time,Frellsen16Topology,Augenstein20Inverse}.
In addition, it was used to design plasmonic antennas using frequency-domain methods\cite{wadbro15topology,christiansen19nonlinear,Zhou21Inverse}.
Christiansen et al.\cite{christiansen19nonlinear} proposed a non-linear interpolation scheme that was successful to enable TopOpt of plasmonic antennas near their surface plasma frequency using the FEM method.

Structural perturbations or changes in material properties, caused for example by fabrication tolerance or temperature variations, raise the demand to account for the broadband performance of optical components\cite{giannini11plasmonic}.
At optical and near-infrared wavelengths, various materials exhibit dispersion, that can be modelled via Drude, Lorentz, or critical-points functions\cite{Taflove05,okoniewaki_simple_97,Prokopidis_Unified_13}. Below their plasma frequency, the Drude model is commonly used to describe the dispersive properties of metals such as silver, gold, and aluminum\cite{vial_improved_2005}. The model also describes the permittivity of conductive oxides, such as indium tin oxide (ITO)\cite{Alam16Large}, and conductive polymers\cite{Karst2021}, including their dielectric and epsilon-near-zero regions.

In this paper, we introduce TopOpt of dispersive optical materials in the time-domain and aim at broadband designs. 
We base our algorithm on the FDTD method\cite{Taflove05} and the Drude model to describe the material dispersion.
To the best of our knowledge, this is the first time the FDTD method is used for TopOpt of plasmonic devices in the time domain.
To present the algorithm, we conduct the optimization of plasmonic nanoantennas, with the goal to maximize the electric energy in a specified region by finding the distribution of the electric permittivity in a design domain.
To evaluate the gradient of the objective function efficiently, we employ the adjoint-field method and provide sensitivity analysis based on Maxwell's equations in their first-order form.
During the optimization, the interpolation of the design material between dielectric and metal enables the surface plasmon frequency $\omega_{sp}$ to develop inside the frequency band of interest\cite{raether88surface,maier07plasmonics}. Plasmonic effects, such as field localization and enhancement, are amplified close to $\omega_{sp}$. This leads to hot-spots within the optimization domain that prevent the algorithm from converging to well performing designs.
To overcome these convergence issues, we exploit the conductivity term in Maxwell's equations to introduce an artificial damping.
This damping counteracts the high-field localization during the optimization process and enables the algorithm to converge to good designs.
The developed algorithm is demonstrated by optimizing 2D (TM and TE) and 3D silver nanoantennas operating near plasma and infrared frequencies.
In all cases, the algorithm produces novel designs with outstanding performance, which demonstrates its potential to offer new opportunities for optimizing dispersive optical nanostructures and metamaterials.

\section{Optimization problem setup} 
\label{S:ProMod}
\begin{figure}
\includegraphics[trim = 0mm 0mm 0mm 3mm, clip ,width=1.0\columnwidth,draft=false]{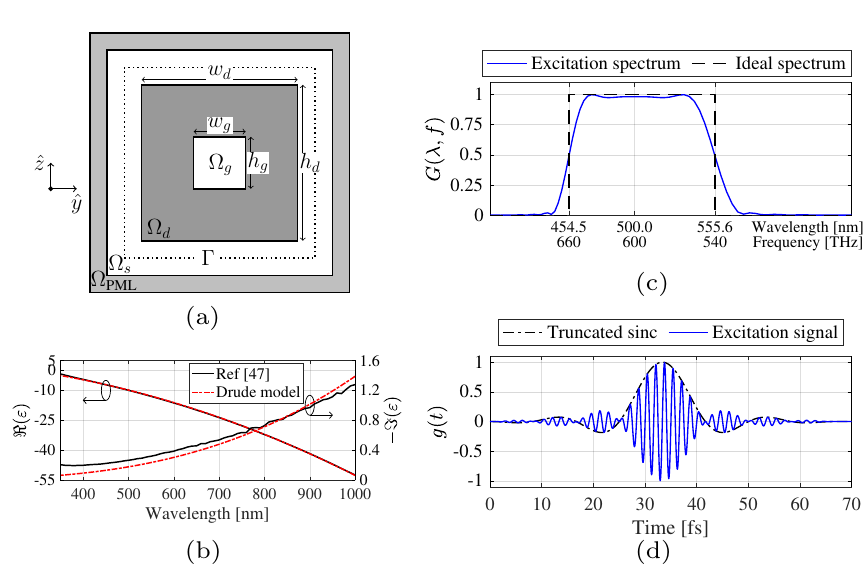}
\caption{(a) Design domain $\Omega_{d}$ where silver is to be distributed to maximize the electric energy at the observation domain $\Omega_g$. $\Omega_{d}$ is contained in a background medium $\Omega_{s}$ and the computational domain is truncated by perfect matched layers $\Omega_\text{PML}$. The boundary $\Gamma$ is used for plane-wave injection. 
(b) Complex permittivity of silver versus wavelength\cite{McPeak15Plasmonic}, and a fitting via the Drude model in the wavelength range $350$--$1000$\,nm. (c) Frequency domain and (d) time domain plots of a sinc signal truncated after four lobes, smoothed using a Hanning window and modulating a carrier signal with a frequency of 600\,THz (i.e., $\lambda_0=500$\,nm).
The modulated signal has a bandwidth of 20\% at half-maximum.}
\label{Geom2D}
\end{figure}

In this section, we present the setup of the optimization problem that will be used to inverse design 2D and 3D nanostructures. 
The computational domain $\Omega$ consists of $\Omega_g \cup \Omega_{d} \cup \Omega_{s} \cup \Omega_\text{PML}$, as shown in Fig.\,\ref{Geom2D}(a). 
The domain $\Omega_g$ is an observation region where the electric energy is to be maximized. We assume that $\Omega_g$ is a dispersionless dielectric medium with relative permittivity $\varepsilon_g$ and has an area $w_g\times h_g$.
Inside the design domain $\Omega_{d}=w_d\times h_d$, we aim to distribute a dispersive material to form the nanoantenna structure. We consider materials with relative permittivity described by the Drude model:
\begin{equation}\label{DrudeSusceptibiltiyF}
\begin{aligned}
\varepsilon^{Drude}(\omega) = \varepsilon_\infty -\frac{\psi}{ \omega^2 -j \omega \gamma_p},
\end{aligned}
\end{equation}
where $\varepsilon_\infty$ is the high-frequency permittivity; $\gamma_p$ is the collision rate; $\psi= \omega_{p}^{2}=\frac{ne^2}{m_e \varepsilon_0}$ is the square of the plasma frequency; $\varepsilon_0$ is the vacuum permittivity; $n$, $m_e$, and $e$ are the electrons' density, effective mass, and charge, respectively. We use the $e^{j\omega t}$ convention.

In this paper, we consider silver in the wavelength range $350$--$1000$\,nm, where we fit its measured complex permittivity, adapted from McPeak et al.\cite{McPeak15Plasmonic}, via a Drude model with parameters $\varepsilon_\infty = 4.469$, $\omega_p = 1.426\times10^{16}$\,rad/s, and $\gamma_p = 4.571\times10^{13}$\,rad/s, as shown in Fig.\,\ref{Geom2D}(b).
The background space $\Omega_{s}$ hosts the design domain $\Omega_d$, and has a constant relative permittivity $\varepsilon_s$. For simplicity, we use $\varepsilon_s=\varepsilon_g=1$, but deviation from these values is also possible.
The computational domain is terminated by perfectly matched layers $\Omega_\text{PML}$.
In our analysis, we use the total-field scattered-field formulation\cite{Taflove05} to inject a plane-wave through the boundary $\Gamma$, which is located in $\Omega_{s}$. 
A similar setup was previously used to optimize TM nanoantennas with frequency-domain methods\cite{wadbro15topology,christiansen19nonlinear}.

To optimize in the time-domain, the spectral content of the excitation signal determines the desired bandwidth of the structure under optimization. 
Ideally, we would like to excite the system with a signal that has a rectangular spectrum (see Fig.\,\ref{Geom2D}(c)). Unfortunately, this corresponds to a sinc signal of infinite duration. Thus, in order to keep the simulation time reasonable, the sinc signal is truncated to a few lobes, as shown in Fig.\,\ref{Geom2D}(d).
Such signal modulates a carrier with a frequency corresponding to the center of the spectral window of interest.
In addition, we use a Hanning window to reduce the ripples in the excitation spectrum.
This processing results in rounding the edges of the spectral window.

We formulate the conceptual optimization problem 
      \begin{equation}\label{OptPro} 
		\begin{aligned}
  			\mathop {\text{maximize}}\limits_{\varepsilon(x) \in [ \varepsilon_s, \varepsilon^{Drude}(\omega)]} \qquad &  W \\
  			\text{  subject to: } 
			&  \text{the governing equations,}\enspace \\
			&  \text{and a specified spectral content,} 
		\end{aligned} 
	\end{equation}	
where 
\begin{equation}\label{ObjectFunc}
\begin{aligned}
W &=   \frac{1}{2}\int_{\Omega_g} \!\int_0^T \varepsilon_g \boldsymbol{\mathcal{E}}^2 dt\,d\Omega 
\end{aligned}
\end{equation}
is the electric energy in $\Omega_g$, $\boldsymbol{\mathcal{E}}$ is the time-dependent electric field, and $T$ is the observation time.
The statement of the optimization problem is to find, for each point $x$ in $\Omega_d$, the permittivity distribution $\varepsilon(x) \in [\varepsilon_s, \varepsilon^{Drude} (\omega)]$ that maximizes the electric energy $W$ in $\Omega_g$ subject to a specified excitation spectrum.
 The domain $\Omega_g$ corresponds to the gap of a plasmonic nanoantenna where the incoming field is to be enhanced, or to a focus region where the energy is collimated by the designed structure. A detailed description of the numerical treatments of the optimization problem, including the sensitivity analysis and the numerical solution, is given in the Methods section.

\subsection{Density-based interpolation and convergence issues}

\begin{figure} 
\includegraphics[trim = 3mm 0mm 2mm 1mm, clip,width=1.0\columnwidth,draft=false]{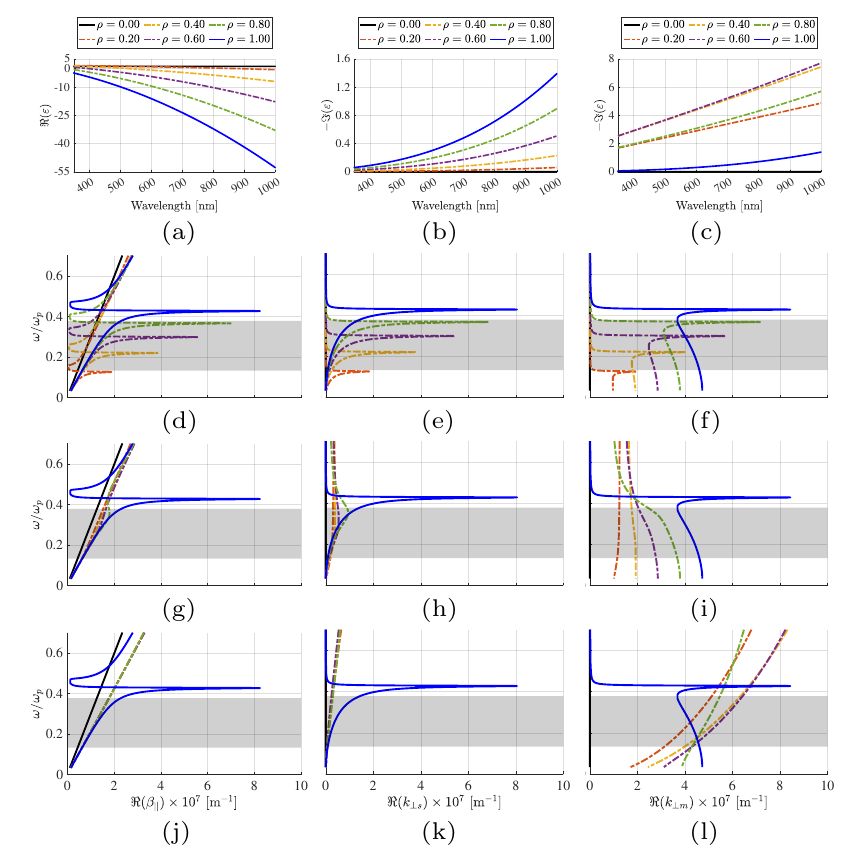}
\caption{Interpolating between the permittivity of free-space ($\rho \!=\! 0$) and silver ($\rho \!=\! 1$) using expression \eqref{DrudeSusceptibiltiyF3}.  (a) Real and (b) Imaginary part of $\varepsilon(\omega,\rho)$ when $\sigma_\text{max} \!=\!0$~S/m. (c) Imaginary part of $\varepsilon(\omega,\rho)$ when $\sigma_\text{max} \!=\!5 \!\times\!10^{5}$~S/m (the real part is the same as (a)). Dispersion diagrams of the surface plasmon polaritons when interpolating between air and silver using (d)-(f) $\sigma_\text{max} \!=\!0$~S/m, (g)-(i) $\sigma_\text{max} \!=\!5 \!\times\!10^{5}$~S/m, and (j)-(l) $\sigma_\text{max} \!=\!5 \!\times\!10^{6}$~S/m. The grey strip marks the spectral window $350$--$1000$\,nm.
}
\label{RhoEpsilonSigma0}
\end{figure}

In density-based topology optimization, we use the permittivity function of the material as our design variable. For each point $x$ in $\Omega_{d}$ we want the permittivity at that point to be either that of the Drude material $\varepsilon^{Drude} (\omega)$ or that of the background space $\varepsilon_s$. In order to interpolate between the background space and the design material, a density variable $\rho_i$ is introduced to describe the material at each edges of the computational grid in $\Omega_d$. The vector $\boldsymbol{\rho} = [\rho_1\, \rho_2\, \cdots \rho_i\, \cdots \rho_M]$ is used to hold the $M$ design variables of the optimization problem.
 Since we aim to use gradient-based methods to solve the topology optimization problem, the entries of the design vector are allowed to attain values between $0$ and $1$ during the optimization process. However, to obtain a manufacturable design, the final density vector must hold only the binary values $0$ or $1$. We map each design variable $\rho_i$ to the physical material parameters using the following interpolation scheme:

\begin{equation}\label{DrudeSusceptibiltiyF3}
\begin{aligned}
 \varepsilon(\omega,\rho_i) = \varepsilon_{\infty\, i} -\frac{\psi_i}{ \omega^2 -j \omega \gamma_p} - j \frac{\sigma_i}{\omega \varepsilon_0},
\end{aligned}
\end{equation}
where
\begin{subequations} \label{mapping}
\begin{align}
\varepsilon_{\infty\,i} &=  \varepsilon_{s} + \rho_i (\varepsilon_\infty - \varepsilon_{s}), \\
\psi_i &=  \psi_s + \rho_i (\psi-\psi_s),   \\
\sigma_i &= \rho_i(1-\rho_i) \sigma_\text{max} \label{mapping:c}
\end{align} 
\end{subequations}
are our three design variables (the dependence on $\rho_i$ is carried by the $i$ subscript). 
The parameters $\psi_i$ and $\varepsilon_{\infty\,i}$ perform a linear interpolation between the physical parameters of the background space $\{\varepsilon_{s},\psi_s\}$ and those of the design material $\{\varepsilon_\infty,\psi\}$. We use $\psi_s = \psi/100$ to ensure $\psi>0$, thus avoiding the singularity in Eq.\,\eqref{DerivativePsiRev}. In fact, the value $\rho_i=1$ corresponds to the Drude model $\varepsilon^{Drude} (\omega)$, and $\rho_i\!=\!0$ sufficiently approximates the background space permittivity $\varepsilon_s$.
To overcome convergence issues, we modify the Drude model to include an artificial conductivity $\sigma_i$, that plays a temporary role only during the optimization. In order to do so, we use a parabolic profile, so that the conductivity is zero for $\rho_i=0$ and $\rho_i=1$, and reaches its maximum value $\sigma_\text{max}$ for $\rho_i \!=\! 0.5$. The value $\sigma_\text{max}$ must be carefully chosen, as discussed later.
Fig.\,\ref{RhoEpsilonSigma0}(a)-(c) show the interpolation between the permittivity of free space ($\rho=0$) and silver ($\rho=1$) using the design permittivity model $\varepsilon(\omega,\rho)$ in Eq. \,\eqref{DrudeSusceptibiltiyF3}, where we drop the index $i$ for brevity.

The presence of metal and dielectric in the design domain leads to localization and enhancement of the electric field due to surface plasmon modes arising during the optimization process. This hinders the convergence of the objective function.
To understand the reason for such convergence problem, we use, as an illustrative model, the dispersion relation of surface plasmon polaritons (SSPs) at a flat interface between a dielectric $\varepsilon_s$ and our design material $\varepsilon(\omega,\rho)$\cite{raether88surface,maier07plasmonics}: 
\begin{subequations} \label{Dispersion}
\begin{align}
\beta_{||}(\omega,\rho) &= k_0 \sqrt{\frac{\varepsilon(\omega,\rho) \varepsilon_s}{\varepsilon(\omega,\rho)+\varepsilon_s}}  \\
k_{\perp s}(\omega,\rho) &= \sqrt{ \beta_{||}^2 -  k_0^2 \varepsilon_s}\\
k_{\perp m}(\omega,\rho) &= \sqrt{ \beta_{||}^2 -  k_0^2 \varepsilon(\omega)}
\end{align} 
\end{subequations}
where $\beta_{||}$ is the propagation constant parallel to the interface; $k_{\perp s}$ and $k_{\perp m}$ are attenuation factors normal to the interface inside the dielectric and metal, respectively; and $k_0 = \omega\sqrt{\mu_0\varepsilon_0}$, as shown in Figs.\,\ref{RhoEpsilonSigma0}(d)-(f) for values of $\rho$ between air ($\rho\!=\!0$) and silver ($\rho\!=\!1$).
The horizontal gray strip marks the wavelength interval of interest $350$--$1000$\,nm.
Inspecting the silver case ($\rho \!=\! 1$), we see that the value of $\beta_{||}$ reaches a maximum close to the surface plasmon frequency $\omega_{sp}$\cite{raether88surface,maier07plasmonics}, which by design is located outside the wavelength window of interest.
The values of $k_{\perp s}$ and $k_{\perp m}$ also reach a maximum close to $\omega_{sp}$, which indicates a large wave attenuation in the directions normal to the interface.
In other words, the wave propagates with the wavenumber $\beta_{||}$, and it is spatially highly localized at the interface region.
The range of frequencies $\omega\!>\!\omega_{sp}$ is not of interest for SPPs. 
For intermediate values of $\rho$, the material permittivity changes from metal to dielectric, and $\omega_\text{sp}$ moves across the wavelength window of interest. This results in moving the peaks of the dispersion curves into the wavelength window of interest, as shown in Figs.\,\ref{RhoEpsilonSigma0}(d)-(f) for $\sigma_\text{max} = 0$~S/m.
The high-field localization, associated with large amplitudes of $k_{\perp s}$, $k_{\perp m}$ and $\beta_{||}$, counteracts the objective function that aims to maximize the electric field in the domain $\Omega_g$. This conflict prevents the algorithm from converging to well-performing designs.

With the aim to reduce the peaks in the dispersion diagrams at intermediate values of $\rho$, we estimate the value of $\sigma_\text{max}$ based on a parameter sweep, as shown in Figs.\,\ref{RhoEpsilonSigma0}(g)-(i) for $\sigma_\text{max} = 5\times10^5$~S/m.
We note that a too-large value of $\sigma_\text{max}$ leads to an intermediate material with conductivity higher than silver, as shown in Fig.\,\ref{RhoEpsilonSigma0}(l) for $\sigma_\text{max} \!=\! 5\times10^6$~S/m, thus causing less wave penetration inside the material.
A too-large value of $\sigma_\text{max}$ prevents the algorithm from converging to black and white designs since the structure becomes less lossy for such a value, while a too-small value is not enough to introduce sufficient damping to counteract the plasmonic effects.

\section{Methods}
\label{Methods}
\subsection{Governing equations and sensitivity analysis}
\label{SenAna}
Inside the computational domain, that we assume source-free and non-magnetic with vacuum permeability $\mu_0$, the time-dependent electric field $\boldsymbol{\mathcal{E}}$ and magnetic field $\boldsymbol{\mathcal{H}}$ are governed by Maxwell's equations
\begin{subequations}\label{MaxwellTime}
\begin{align}
 \partial_t \boldsymbol{\mathcal{D}} - \nabla\times \boldsymbol{\mathcal{H}} = \bs 0\label{MaxwellTime:amp}, \\
\mu_0 \partial_t \boldsymbol{\mathcal{H}}+ \nabla\times\boldsymbol{\mathcal{E}} = \bs 0,  
\end{align}
\end{subequations}
where the electric displacement field $\boldsymbol{\mathcal{D}}$ models the optical response of the materials through the frequency domain relation
\begin{equation}\label{EFlux}
\begin{aligned}
 \bs D(\omega) & = \varepsilon_{0} \varepsilon(\omega) \bs E(\omega). 
\end{aligned}
\end{equation}
In our case, we use the design permittivity model
\begin{equation}\label{DrudeSusceptibiltiyF2}
\begin{aligned}
 \varepsilon(\omega) = \varepsilon_\infty -\frac{\psi}{ \omega^2 -j \omega \gamma_p} - j \frac{\sigma}{\omega \varepsilon_0},
\end{aligned}
\end{equation}
that consists of a Drude model and an additional artificial conductivity term that operates only during the optimization process to guarantee convergence.

By substituting Eq.\eqref{DrudeSusceptibiltiyF2} and Eq.\eqref{EFlux} into Eq.\eqref{MaxwellTime:amp}, we rewrite the time-domain Maxwell's equation \eqref{MaxwellTime} as
\begin{subequations}\label{MaxwellTime2}
\begin{align}
 \varepsilon_0 \varepsilon_\infty \partial_t \boldsymbol{\mathcal{E}} + \boldsymbol{\mathcal{J}} + \sigma \boldsymbol{\mathcal{E}} - \nabla\times \boldsymbol{\mathcal{H}} = \bs 0,\\
\partial_t \boldsymbol{\mathcal{J}} + \gamma_p \boldsymbol{\mathcal{J}} - \varepsilon_{0} \psi \boldsymbol{\mathcal{E}} = \bs 0, \label{MaxwellTime2:b}\\
  \mu_0 \partial_t \boldsymbol{\mathcal{H}} + \nabla\times\boldsymbol{\mathcal{E}}= \bs 0, 
\end{align}
\end{subequations}
where the dispersion of the medium is embedded in the polarization current density $\boldsymbol{\mathcal{J}}$, with Eq.\eqref{MaxwellTime2:b} being the time-domain equivalent of
\begin{equation}\label{Polarization}
\begin{aligned}
\boldsymbol{J}(\omega) &=  \frac{\varepsilon_{0} \psi}{ j \omega + \gamma_p} \boldsymbol{E}(\omega).
\end{aligned}
\end{equation}

We solve the design problem by using a gradient-based optimization method. 
To do so, we need derivatives of the objective function with respect to the design variables $\varepsilon_\infty$, $\psi$, and $\sigma$.
In this section, we use the adjoint-field method to derive expressions for such derivatives.  

Assuming that the design variables are perturbed by $\delta\varepsilon_\infty$, $\delta\psi$, and $\delta\sigma$, the corresponding first variations of $W$ are 
\begin{subequations}\label{PerturbationII}
\begin{align}
 &\delta_{\varepsilon_\infty} W  = \int_{\Omega_g} \!\int_0^T \varepsilon_g \boldsymbol{\mathcal{E}}\, \delta_{\varepsilon_\infty} \boldsymbol{\mathcal{E}} \,dt\,d\Omega, \\ 
& \delta_\psi W = \int_{\Omega_g} \!\int_0^T \varepsilon_g \boldsymbol{\mathcal{E}}\, \delta_\psi \boldsymbol{\mathcal{E}} \,dt\,d\Omega, \\
&\delta_{\sigma} W = \int_{\Omega_g} \!\int_0^T \varepsilon_g \boldsymbol{\mathcal{E}}\, \delta_{\sigma} \boldsymbol{\mathcal{E}} \,dt\,d\Omega.
\end{align}
\end{subequations}
To find explicit expressions for \eqref{PerturbationII}, we use the system governing equations \eqref{MaxwellTime2} and employ the adjoint-field method\cite{zhang_wideband_14,Hassan15Time}.
In the following, we derive an explicit relation for $\delta_\psi\! W$.
To simplify the derivation, we drop the domain $\Omega_\text{PML}$ and consider $\Gamma$ as the external boundary of the analysis domain $\Omega$, see Fig.\,\ref{Geom2D}(a). 
In addition, we drop the differential $dt$ and $d\Omega$ since they can be inferred from the limits of the integrals.
We consider the initial-boundary-value-problem, 
\begin{subequations}\label{IBVP}
\begin{align}
\varepsilon_0 \varepsilon_\infty \partial_t \boldsymbol{\mathcal{E}} +\boldsymbol{\mathcal{J}} + \sigma \boldsymbol{\mathcal{E}} - \nabla\times \boldsymbol{\mathcal{H}} = \bs 0  \phantom{{}^{+}} & \,\,\,\text{ in } \Omega, t>0 \\
\partial_t \boldsymbol{\mathcal{J}} + \gamma_p \boldsymbol{\mathcal{J}} - \varepsilon_{0} \psi\boldsymbol{\mathcal{E}} = \bs 0  \phantom{{}^{+}}& \,\,\, \text{ in } \Omega, t>0 \\
\mu_0 \partial_t \boldsymbol{\mathcal{H}} + \nabla\times\boldsymbol{\mathcal{E}}= \bs 0 \phantom{{}^{+}}&  \,\,\,\text{ in } \Omega, t>0 \\ 
\boldsymbol{\mathcal{E}}_{t} + \eta \,\boldsymbol{n} \times\boldsymbol{\mathcal{H}} =g  \phantom{{}^{+}} & \,\,\, \text{ on }  \Gamma, t>0 \label{IBVPBC}\\
\boldsymbol{\mathcal{E}}= \bs 0, \boldsymbol{\mathcal{J}} = \bs 0  \phantom{{}^{+}},  \boldsymbol{\mathcal{H}} = \bs 0  \phantom{{}^{+}}&\,\,\,\text{ in } \Omega, t = 0,
\end{align}
\end{subequations}
where $\eta=\sqrt{\mu_0/(\varepsilon_0\varepsilon_s)}$ is the intrinsic impedance of the domain $\Omega_{s}$, and $\boldsymbol{\mathcal{E}}_{t} = \boldsymbol{\mathcal{E}}-\boldsymbol{n}(\boldsymbol{\mathcal{E}}\cdot \boldsymbol{n})$ with $\boldsymbol{n}$ denoting the outward unit normal at $\Gamma$. 
The boundary condition \eqref{IBVPBC} is used to impose an incoming excitation $g$ through $\Gamma$.
To simplify the notation, the symbol $\delta$ is temporarily used to denote the perturbation of the fields with respect to $\psi$.
We differentiate the system of equations \eqref{IBVP} with respect to $\psi$, 
\begin{subequations}\label{dIBVP}
\begin{align}
\varepsilon_0 \varepsilon_\infty \partial_t \delta\boldsymbol{\mathcal{E}} +\delta\boldsymbol{\mathcal{J}} + \sigma \delta\boldsymbol{\mathcal{E}} - \nabla\times \delta\boldsymbol{\mathcal{H}} = \bs 0 & \,\,\,\text{in } \Omega, t>0 \label{dIBVPa}\\
\partial_t \delta\boldsymbol{\mathcal{J}} + \gamma_p \delta\boldsymbol{\mathcal{J}} - \varepsilon_{0} \psi \delta\boldsymbol{\mathcal{E}}-\varepsilon_{0} \boldsymbol{\mathcal{E}} \delta\psi = \bs 0 &\,\,\, \text{in } \Omega, t>0 \label{dIBVPb} \\
\mu_0 \partial_t \delta\boldsymbol{\mathcal{H}} + \nabla\times\delta\boldsymbol{\mathcal{E}}= \bs 0 & \,\,\,\text{in } \Omega, t>0 \label{dIBVPc}\\ 
\delta\boldsymbol{\mathcal{E}}_{t} + \eta \,\boldsymbol{n} \times\delta\boldsymbol{\mathcal{H}} =\bs 0 & \,\,\,\text{on } \Gamma, t>0 \label{dIBVPd}\\
\delta\boldsymbol{\mathcal{E}} = \bs 0, \delta\boldsymbol{\mathcal{J}} = \bs 0, \delta\boldsymbol{\mathcal{H}} = \bs 0 &\,\,\,\text{in } \Omega, t =0.
\end{align}
\end{subequations}

We define the adjoint fields $\boldsymbol{\mathcal{E}}^{*}$, $\boldsymbol{\mathcal{J}}^{*}$, and $\boldsymbol{\mathcal{H}}^{*}$.
We perform the scalar product of \eqref{dIBVPa}, \eqref{dIBVPb}, and \eqref{dIBVPc} with $\boldsymbol{\mathcal{E}}^{*}$, $\frac{\boldsymbol{\mathcal{J}}^{*}}{\varepsilon_0 \psi}$, and $\boldsymbol{\mathcal{H}}^{*}$, respectively.
We add the result of multiplication, integrating over the whole analysis domain $\Omega$ and the observation interval $(0,T)$, and applying integration by parts, we obtain 
\begin{flalign}\label{IntIBVP}
& \varepsilon_0  \varepsilon_\infty \,  \boldsymbol{\mathcal{E}}^{*} \, \delta\boldsymbol{\mathcal{E}} \big|_0^T -  \int_\Omega\int_0^T \varepsilon_0  \varepsilon_\infty \partial_t \, \boldsymbol{\mathcal{E}}^{*} \,    \delta \boldsymbol{\mathcal{E}}+ \int_\Omega\int_0^T  \boldsymbol{\mathcal{E}}^{*} \, \delta \boldsymbol{\mathcal{J}}  +\nonumber\\
 & \int_\Omega\int_0^T  \sigma \boldsymbol{\mathcal{E}}^{*} \,   \delta \boldsymbol{\mathcal{E}}  
 -\!\int_{\Gamma} \int_0^T (\boldsymbol{n}\times \delta \boldsymbol{\mathcal{H}}) \, \boldsymbol{\mathcal{E}}^{*} - \!\int_{\Omega}\int_0^T (\nabla \times \boldsymbol{\mathcal{E}}^{*}) \, \delta \boldsymbol{\mathcal{H}} \nonumber\\
& - \frac{\boldsymbol{\mathcal{J}}^{*}}{\varepsilon_0 \psi} \, \delta\boldsymbol{\mathcal{J}}  \big|_0^T+ \int_\Omega\int_0^T \partial_t \, \boldsymbol{\mathcal{J}}^{*} \, \frac{\delta \boldsymbol{\mathcal{J}}}{\varepsilon_0 \psi} - \int_\Omega\int_0^T  \gamma_p \boldsymbol{\mathcal{J}}^{*} \, \frac{\delta \boldsymbol{\mathcal{J}}}{\varepsilon_0 \psi} + \nonumber\\ 
& \int_\Omega\int_0^T \boldsymbol{\mathcal{J}}^{*} \,   \delta \boldsymbol{\mathcal{E}}  +\frac{\boldsymbol{\mathcal{E}}  \, \boldsymbol{\mathcal{J}}^{*}}{\psi}  \delta \psi + \mu_0 \boldsymbol{\mathcal{H}}^{*} \, \delta \boldsymbol{\mathcal{H}}\big|_0^T  - \nonumber \\ 
& \int_\Omega\int_0^T \mu_0 \partial_t  \boldsymbol{\mathcal{H}}^{*} \, \delta \boldsymbol{\mathcal{H}} +\int_{\Gamma}\int_0^T (\boldsymbol{n}\times \delta \boldsymbol{\mathcal{E}}) \, \boldsymbol{\mathcal{H}}^{*}  \nonumber \\ 
&+ \int_{\Omega}\int_0^T \nabla\times \boldsymbol{\mathcal{H}}^{*} \, \delta \boldsymbol{\mathcal{E}}  = \bs 0.  
\end{flalign}
We assume that the adjoint fields satisfy the terminal conditions $\boldsymbol{\mathcal{E}}^{*}\!=\!\boldsymbol{\mathcal{J}}^{*}\!=\! \boldsymbol{\mathcal{H}}^{*}\!=\!  \bs 0$ at $t\!=\!T$. 
By arranging the terms in \eqref{IntIBVP}, utilizing \eqref{dIBVPd}, and adding and subtracting $\delta_\psi W$, we obtain
\begin{flalign}\label{IntIBVPR}
&\int_{\Omega}\int_0^T (-\varepsilon_0  \varepsilon_\infty \partial_t \, \boldsymbol{\mathcal{E}}^{*} +   \boldsymbol{\mathcal{J}}^{*}+ \sigma \boldsymbol{\mathcal{E}}^{*}  + \nabla\times \boldsymbol{\mathcal{H}}^{*})  \,  \delta \boldsymbol{\mathcal{E}}  \nonumber  \\
&+ \int_{\Omega}\int_0^T (\partial_t \, \boldsymbol{\mathcal{J}}^{*} - \gamma_p \boldsymbol{\mathcal{J}}^{*} + \varepsilon_0 \psi  \boldsymbol{\mathcal{E}}^{*}) \, \frac{\delta \boldsymbol{\mathcal{J}}}{\varepsilon_0 \psi}  \nonumber  \\
&+\int_{\Omega}\int_0^T (-\partial_t  \mu \, \boldsymbol{\mathcal{H}}^{*} -\nabla\times \boldsymbol{\mathcal{E}}^{*} )\, \delta \boldsymbol{\mathcal{H}}  \nonumber \\
&+\int_{\Gamma}\int_0^T (\boldsymbol{\mathcal{E}}^{*}_t - \eta\, \boldsymbol{n}\times \boldsymbol{\mathcal{H}}^{*}) \, \delta \boldsymbol{\mathcal{H}}  \nonumber \\
&+\int_{\Omega}\int_0^T \frac{\boldsymbol{\mathcal{E}} \, \boldsymbol{\mathcal{J}}^{*}}{\psi} \delta \psi  + \delta_\psi W -\int_{\Omega_g} \int_0^T \varepsilon_g \boldsymbol{\mathcal{E}} \, \delta \boldsymbol{\mathcal{E}}   = \bs 0.  
\end{flalign}
If we require 
\begin{subequations}\label{AdjointPsi}
\begin{align}
 -\varepsilon_0  \varepsilon_\infty \partial_t \, \boldsymbol{\mathcal{E}}^{*}  \!\!+\!  \boldsymbol{\mathcal{J}}^{*}\!\!+\! \sigma \boldsymbol{\mathcal{E}}^{*} \!\!+\! \nabla\!\times\! \boldsymbol{\mathcal{H}}^{*}  \!=\!  \varepsilon_g\boldsymbol{\mathcal{E}}  & \quad\text{in } \Omega, t>0 \\
\partial_t \, \boldsymbol{\mathcal{J}}^{*} - \gamma_p \boldsymbol{\mathcal{J}}^{*} \!+\! \varepsilon_0 \psi \boldsymbol{\mathcal{E}}^{*} \!=\! \phantom{\varepsilon}\bs 0  \phantom{\varepsilon}  & \quad\text{in } \Omega, t>0\\
 \partial_t  \mu \, \boldsymbol{\mathcal{H}}^{*} \!+\!\nabla\times \boldsymbol{\mathcal{E}}^{*} \! = \! \phantom{\varepsilon} \bs 0 \phantom{\varepsilon}  & \quad\text{in } \Omega, t>0\\
 \boldsymbol{\mathcal{E}}^{*}_t \!-\! \eta\, \boldsymbol{n}\times \boldsymbol{\mathcal{H}}^{*} \!=\! \phantom{\varepsilon}\bs 0  \phantom{\varepsilon} & \quad\text{on } \Gamma, t>0\\
\boldsymbol{\mathcal{E}}^{*} \!=\! \bs 0, \boldsymbol{\mathcal{J}}^{*} \!=\! \bs 0,  \boldsymbol{\mathcal{H}}^{*} \!=\! \phantom{\varepsilon}\bs 0 \phantom{\varepsilon} & \quad\text{in } \Omega, t \!=\!T,
\end{align}
\end{subequations}
then equation \eqref{IntIBVPR} reduces to
\begin{flalign}\label{DerivativePsi}
\delta_\psi W    = -\int_{\Omega}\int_0^T \frac{\boldsymbol{\mathcal{E}} \, \boldsymbol{\mathcal{J}}^{*}}{\psi}  \delta \psi.
\end{flalign}
Expression \eqref{DerivativePsi} is the directional derivative of $W$ when $\psi$ is perturbed by $\delta\psi$. 
The  gradient of $W$ with respect to $\psi$ can be identified as the integral kernel 
\begin{flalign}\label{GateauxPsi}
\nabla_\psi W     = -\int_0^T \frac{\boldsymbol{\mathcal{E}}\, \boldsymbol{\mathcal{J}}^{*}}{\psi} .
\end{flalign}

The adjoint system \eqref{AdjointPsi} is a terminal-value-problem which, by changing the time variable (i.e., $t\!=\!T-\tau$) and the sign of the magnetic field $\boldsymbol{\mathcal{H}}^{*}$(i.e., to preserve the direction of the Poynting vector), can be written as
\begin{subequations}\label{AdjointPsiRev}
\begin{align}
\varepsilon_0  \varepsilon_\infty \partial_\tau \boldsymbol{\mathcal{E}}^{*}  \!\!+\!  \boldsymbol{\mathcal{J}}^{*}\!\!+\! \sigma \boldsymbol{\mathcal{E}}^{*} \!\!-\! \nabla\!\times\! \boldsymbol{\mathcal{H}}^{*}  \!=\! \varepsilon_g\overleftarrow{\boldsymbol{\mathcal{E}}} & \quad\text{in } \Omega, \tau<T \\
\partial_\tau \boldsymbol{\mathcal{J}}^{*} + \gamma_p \boldsymbol{\mathcal{J}}^{*} - \varepsilon_0 \psi \boldsymbol{\mathcal{E}}^{*} = \bs 0   \phantom{E}\,\, & \quad\text{in } \Omega, \tau<T \\
\partial_\tau  \mu \boldsymbol{\mathcal{H}}^{*} +\nabla\times \boldsymbol{\mathcal{E}}^{*}  = \bs 0   \phantom{E}\,\,  &\quad\text{in } \Omega, \tau<T \\
\boldsymbol{\mathcal{E}}^{*}_t + \eta\, \boldsymbol{n}\times \boldsymbol{\mathcal{H}}^{*} = \bs 0 \phantom{E}\,\, &\quad \text{on } \Gamma, \tau<T \\
\boldsymbol{\mathcal{E}}^{*} \!=\! \bs 0, \boldsymbol{\mathcal{J}}^{*} \!=\! \bs 0,  \boldsymbol{\mathcal{H}}^{*} = \bs 0 \phantom{E}\,\, & \quad\text{in } \Omega, \tau = 0, 
\end{align}
\end{subequations}
and \eqref{GateauxPsi} becomes
\begin{flalign}\label{DerivativePsiRev}
\nabla_\psi W    = -\int_0^T \frac{\overleftarrow{\boldsymbol{\mathcal{E}}} \, \boldsymbol{\mathcal{J}}^{*}}{\psi},
\end{flalign}
where $\overleftarrow{\boldsymbol{\mathcal{E}}} = \boldsymbol{\mathcal{E}}(T-\tau)$ is the electric field of the forward system \eqref{IBVP} reversed in time. 
The singularity of \eqref{DerivativePsiRev} can be avoided by ensuring the condition  $\psi=\omega_{p}^2>0$.
The only difference between the adjoint system \eqref{AdjointPsiRev} and the forward system \eqref{IBVP} is the source. 
In the forward system \eqref{IBVP}, the source is  a plane-wave imposed through the boundary $\Gamma$, see \eqref{IBVPBC}.
In the adjoint system \eqref{AdjointPsiRev}, the source is the time-reversal of the forward electric field monitored at $\Omega_g$.

By differentiating system \eqref{IBVP} with respect to $\varepsilon_\infty$ and $\sigma$, and following similar procedures as before, we obtain the same adjoint system \eqref{AdjointPsiRev} and the following gradient expressions 
\begin{flalign}\label{DerivativeEps}
\nabla_{\varepsilon_\infty} W    = -\int_0^T\varepsilon_0 \overleftarrow{\boldsymbol{\mathcal{E}}} \, \partial_\tau \boldsymbol{\mathcal{E}}^{*},
\end{flalign}
and 
\begin{flalign}\label{DerivativeSig}
\nabla_{\sigma} W  = -\int_0^T \overleftarrow{\boldsymbol{\mathcal{E}}} \, \boldsymbol{\mathcal{E}}^{*}.
\end{flalign}
Therefore, to evaluate the gradient components of the objective function, we solve the forward system \eqref{IBVP} and the adjoint system \eqref{AdjointPsiRev}, then we use \eqref{DerivativePsiRev}, \eqref{DerivativeEps}, and \eqref{DerivativeSig} to form the full gradient components.

\subsection{Numerical treatments and optimization algorithm}
\label{S:OptAlg}
\label{S:NumTre}

We solve numerically the system of governing equations, discussed in the previous section, using the FDTD method\cite{Taflove05}.
We adopt the auxiliary differential equation approach to implement the Drude model in the FDTD method, and the uniaxial perfectly matched layer (UPML) is used to simulate the open-space radiation boundary condition\cite{okoniewaki_simple_97,Taflove05}.
The computational domain is discretized into uniform square (2D) or cubical (3D)\, Yee cells with spatial steps $\Delta x \!=\! \Delta y \!=\! \Delta z$ in all Cartesian directions, and the total-field scattered-field approach is used to impose the plane-wave excitation.

In the forward system, we use the FDTD method and discretize the electric field at full-time indices and the magnetic field at half-time indices.
In the adjoint system, however, the FDTD discretization of the electric- and magnetic fields are performed at half-time indices and full-time indices, respectively\cite{Hassan14Topology,Hassan15Time}.
The discretized objective function is 
\begin{equation}\label{ObjectFuncDis}
\begin{aligned}
\tilde{W} &=   \frac{ \varepsilon_g \Delta t  (\Delta x)^3}{2} \sum_{\tilde{\Omega}_g} \sum_{n=0}^{N}  (\boldsymbol{\mathcal{\tilde{E}}}^{n})^2  
\end{aligned}
\end{equation}
where $\boldsymbol{\mathcal{\tilde{E}}}^{n}$ is the discretized electric field at time index $n$, $\Delta t$ is the FDTD's temporal discretization step, and $N$ is the number of time steps used in the simulations. 
Based on the FDTD discretization of the forward and adjoint systems, the pointwise derivatives of the gradient expressions given in Section\,\ref{SenAna}, are 
\begin{subequations}\label{dObjectFunc}
\begin{align}
\frac{\partial \tilde{W}}{\partial \psi_i} &=   -\frac{ \Delta t  (\Delta x)^3}{\psi_i} \sum_{n=0}^{N}  
\boldsymbol{\mathcal{\tilde{E}}}_{i}^{N-n} \frac{ \boldsymbol{\mathcal{\tilde{J}}}_{i}^{*n+\frac{1}{2}}  + \boldsymbol{\mathcal{\tilde{J}}}_{i}^{*n-\frac{1}{2}} }{2} \label{dpsiObjectFunc}\\
\frac{\partial \tilde{W}}{\partial \varepsilon_{\infty\,i}} &=   - \varepsilon_0 (\Delta x)^3 \sum_{n=0}^{N}  
\boldsymbol{\mathcal{\tilde{E}}}_{i}^{N-n} (\boldsymbol{\mathcal{\tilde{E}}}_{i}^{*n+\frac{1}{2}} - \boldsymbol{\mathcal{\tilde{E}}}_{i}^{*n-\frac{1}{2}}) \label{depsObjectFunc}\\
\frac{\partial \tilde{W}}{\partial \sigma_{i}} &=    -  \Delta t  (\Delta x)^3  \sum_{n=0}^{N}  
\boldsymbol{\mathcal{\tilde{E}}}^{N-n}_i \frac{ \boldsymbol{\mathcal{\tilde{E}}}_{i}^{*n+\frac{1}{2}}  + \boldsymbol{\mathcal{\tilde{E}}}_{i}^{*n-\frac{1}{2}} }{2} \label{dsigmaObjectFunc}
\end{align}
\end{subequations}
where $i$ denotes the index of the $i^{\text{th}}$edge in the design domain.
Note that the temporal averaging of the discrete adjoint fields in \eqref{dpsiObjectFunc} and \eqref{dsigmaObjectFunc} is related to the time shift between the forward and the adjoint discrete systems\cite{Hassan15Time}.

\begin{figure}
\begin{center}
\includegraphics[trim = 0mm 5mm 0mm 0mm, clip,width=0.9\columnwidth,draft=false]{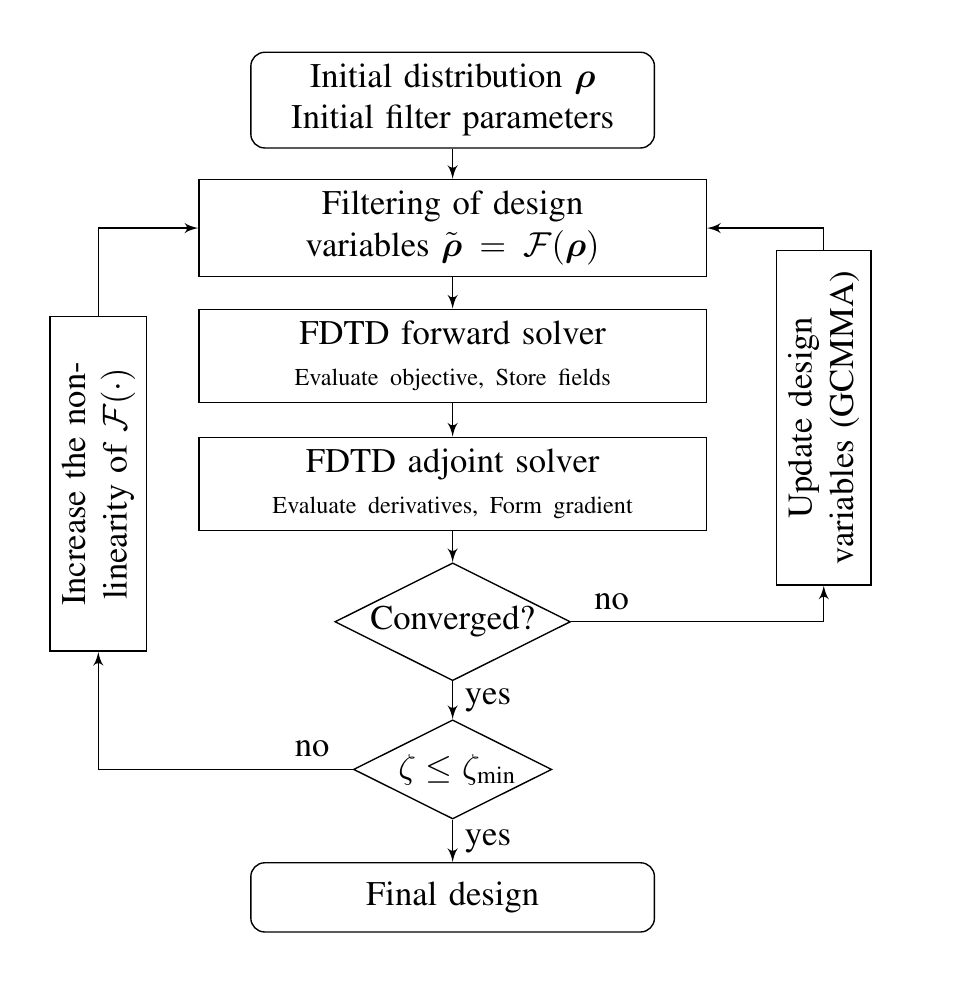}
\end{center}
\caption{Flowchart of the optimization algorithm.}
\label{OptAlg}
\end{figure}

To avoid mesh-dependency or self-penalization issues, in density-based topology optimization it is common to filter the design variables\cite{Borr01,Si07,SvSv13,Hassan14patch,HaWa16}.
That is, instead of using $\rho_i$ in \eqref{mapping}, we replace it with $\tilde{\rho}_i$, where the filtered design vector $\boldsymbol{\tilde{\rho}}$ is obtained through the mapping 
\begin{equation}\label{Filter} 
\begin{aligned}
\boldsymbol{\tilde{\rho}} = \mathcal{F}(\boldsymbol{\rho}).
\end{aligned} 
\end{equation}	
In this work, we use an open-close, nonlinear filter operator $\mathcal{F}(\cdot)$ that consists of a cascade of four $fW$-mean filters\cite{HaWa16}.
The filter has two tuning parameters that determine its size and the level of nonlinearity.
Here, we fix the filter size to a constant value of $5\Delta x$ and only employ the nonlinearity parameter to smoothly decrease the level of greyness in the design during the optimization process.
More details about nonlinear filters and their use in topology optimization can be found in the literature\cite{HaWa16,Hassan2018}.
Using the chain rule, the derivative of the discrete objective function with respect to the design variable $\rho_i$ is evaluated by 
\begin{equation}\label{DiscDerivatives} 
\begin{aligned}
\frac{\partial \tilde{W}}{\partial \rho_i} = \frac{\partial \tilde{\rho}_i}{\partial \rho_{i}}  \frac{\partial \psi_i }{\partial \tilde{\rho}_i}  \frac{\partial \tilde{W}}{\partial \psi_{i}}\!+\!\frac{\partial \tilde{\rho}_i}{\partial \rho_{i}}  \frac{\partial \varepsilon_{\infty\,i}}{\partial \tilde{\rho}_i} \frac{\partial \tilde{W}}{\partial \varepsilon_{\infty\,i}}\!+\!\frac{\partial \tilde{\rho}_i}{\partial \rho_{i}} \frac{\partial \sigma_{i}}{\partial \tilde{\rho}_i} \frac{\partial \tilde{W}}{\partial \sigma_{i}}.
\end{aligned} 
\end{equation}	
We compared the derivatives computed using expression \eqref{DiscDerivatives} against those evaluated by finite differences.
The comparison showed more than $4$\,digits match in precision between the two methods.
We write the discrete version of the optimization problem as
      \begin{equation}\label{OptProDisc} 
		\begin{aligned}
  			\mathop \text{maximize} \limits_{\boldsymbol{\rho}} \qquad &  \tilde{W }\\
  			\text{  subject to: } 			
			&  \text{the governing equations},\enspace \\
			&  \text{a specified spectral content}, \\
			&  0 < \rho_{i} <1,
		\end{aligned} 
	\end{equation}	
which we solve iteratively through the solution of a sequence of subproblems.
Fig.\,\ref{OptAlg} shows the flowchart of the optimization algorithm that we use to solve \eqref{OptProDisc}.
To update the design variables, we use the globally convergent method of moving asymptotes (GCMMA)\cite{SvanbergGlobally}. 
As a stopping criterion for the inner iteration loop, we monitor the norm of the first-order optimality condition after $12$ iterations.
Then, we mark the decrease of this norm by $70$\% as the termination condition of the subproblem.
For the termination of the outer loop, we monitor the decrease of the level of non-discreteness, $\zeta= 4 \tilde{\boldsymbol{\rho}}^T (\boldsymbol{1} -\tilde{\boldsymbol{\rho}})/M$ with $\boldsymbol{1}$ denoting a vector of a length $M$ and all entries equal one\cite{Si07}.
We terminate the optimization process either when the value of $\zeta$ decreases below $\zeta_\text{min}=0.5\%$ or a maximum number of $600$\,iterations is reached.
Then, the entries of the obtained design are thresholded around $\rho_\text{th}=0.5$ to yield the final design.

\section{Results}
\label{S:Res}
In this section, we demonstrate the capabilities of the time-domain optimization method through several design examples of plasmonic nanostructures in 2D (TM and TE) and 3D. 
To enable fast simulation, we implement the FDTD method to execute on graphics processing units (GPUs).
The computations are carried out on nodes equipped with NVidia V100 GPUs and 64\,GB of memory.
Based on the problem size, one call to the Maxwell solver uses a simulation time between less than a minute and a few minutes.
\subsection{2D, TM nanoantennas}

\begin{figure}
\includegraphics[trim = 2mm 0mm 0mm 3mm, clip ,width=1.0\columnwidth,draft=false]{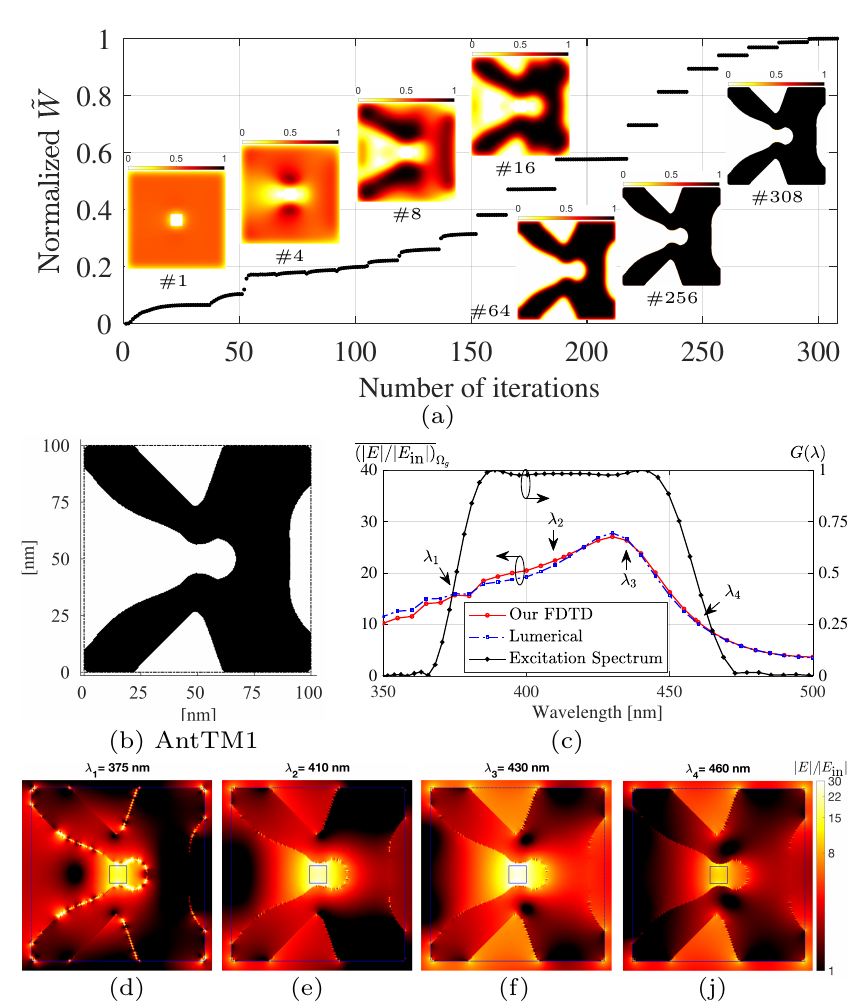}
\caption{(a) Progress of the objective function and some samples showing the development of AntTM1. (b) Topology of AntTM1. (c) Average electric field enhancement in $\Omega_g$ together with the spectrum of the excitation signal. (d)-(j) Field distribution of AntTM1 at $\lambda=375$, $410$, $430$, and $460$\,nm.}
\label{Objective2DEx1_1IniResII}
\end{figure}

We chose a design domain $\Omega_d$ with dimensions $w_d \!=\! h_d\!=\!100\,\text{nm}$, and the observation domain $\Omega_g$ is centered within $\Omega_d$ and has dimensions $w_g \!=\! h_g\!=\!10 \,\text{nm}$, see Fig.\,\ref{Geom2D}(a).
We use a space-step $\Delta x=0.5\,\text{nm}$, and a time-step $\Delta t$ satisfying the Courant stability criterion. 
The simulation domain is truncated by 15\,UPML cells placed 30\,cells away from $\Omega_d$.
The excitation is an $\mathcal{H}_{x}$ polarized plane-wave propagating towards the positive $y$\,axis.
The excitation spectrum has a bandwidth of 20\% at half-maximum and is centered at $413$\,nm, as shown in Fig.\,\ref{Objective2DEx1_1IniResII}(c) along with the results of optimization.
The design variables are mapped to the in-plane permittivity components associated with the Yee edges where the electric field components $\mathcal{E}_y$ and $\mathcal{E}_z$ are located.
Excluding the observation region, the design domain includes 79\,560 design variables (edges).

Figure\,\ref{Objective2DEx1_1IniResII}(a) shows the progress of the normalized objective function $\tilde{W}$ (see Eq. \eqref{ObjectFuncDis} in the Methods section) versus the iteration numbers.
We start the algorithm with a uniform initial distribution $\rho_i \!= \!0.5$ for all design variables.
Included in the same figure are some snapshots to show the development of the design.
The black color indicates silver ($\rho\! =\!1$) and the white color indicates air ($\rho \!=\!0$).
Notably, the main topology of the antenna evolves after only a few tens of iterations.
However, most of the late iterations are used to remove the intermediate material and form crisp boundaries, while the objective function keeps increasing monotonically.
The increase of the objective function could be inferred from Fig.\,\ref{RhoEpsilonSigma0}(i) as follows.
The material interpolation between air and silver allows waves to penetrate deeper into materials with intermediate densities, and therefore, the energy losses inside the antenna structure increases.
As the amount of intermediate material decreases, waves penetrate less inside the device structure which allows the objective function to increase.
The design algorithm converged after $308$\,iterations to a design with a grayness level $\zeta < 0.2$\% (see Methods section for the definition of $\zeta$).
We threshold this design around $\rho_\text{th}=0.5$ and show the final design, AntTM1, in Fig.\,\ref{Objective2DEx1_1IniResII}(b). 
On the side facing the incident wave, the topology of AntTM1 developed as a flared horn, backed by a small cavity region.
On the other side, we see two slightly tilted vertical arms.
This optimized topology shares similarities with results reported in the literature using the FEM method\cite{christiansen19nonlinear}.

Fig.\,\ref{Objective2DEx1_1IniResII}(c) shows the average field enhancement of AntTM1 at the observation domain $\Omega_g$, which correlates well with the spectrum of the excitation signal shown in the same figure.
In the same figure, we cross-validate our computations with the commercial software package Ansys Lumerical FDTD\cite{Ansys}.
Slight differences between the two computations are attributed to differences in geometry descriptions.
Inside $\Omega_g$ and within the main window of the excitation, AntTM1 exhibits more than $20$-fold field enhancement compared to the incident wave.
The peak of the performance, $(\overline{|E|/|{E}_\text{in}|})_{\Omega_g} \!\!=\! 27.8$, occurs at the wavelength $\lambda\!=\!430$\,nm, which resides at long wavelengths in the excitation window.
Figs.\,\ref{Objective2DEx1_1IniResII}(d)-(j) show the electric field distribution of AntTM1 at four wavelengths, marked in Fig.\,\ref{Objective2DEx1_1IniResII}(c).
The electric field is maximum at the observation domain $\Omega_g$, marked by the box at the center.
However, we observe a field localization and enhancement at the device's boundaries for short wavelengths, which justifies the decrease of the energy enhancement in $\Omega_g$ at short wavelengths.
We attempt to improve the optimization results further by exploring three investigations in the following paragraphs.
\subsubsection{Effect of the design domain size}
\begin{figure}
\vspace{-10pt}
\includegraphics[trim = 0mm 0mm 0mm 0mm, clip ,width=1.0\columnwidth,draft=false]{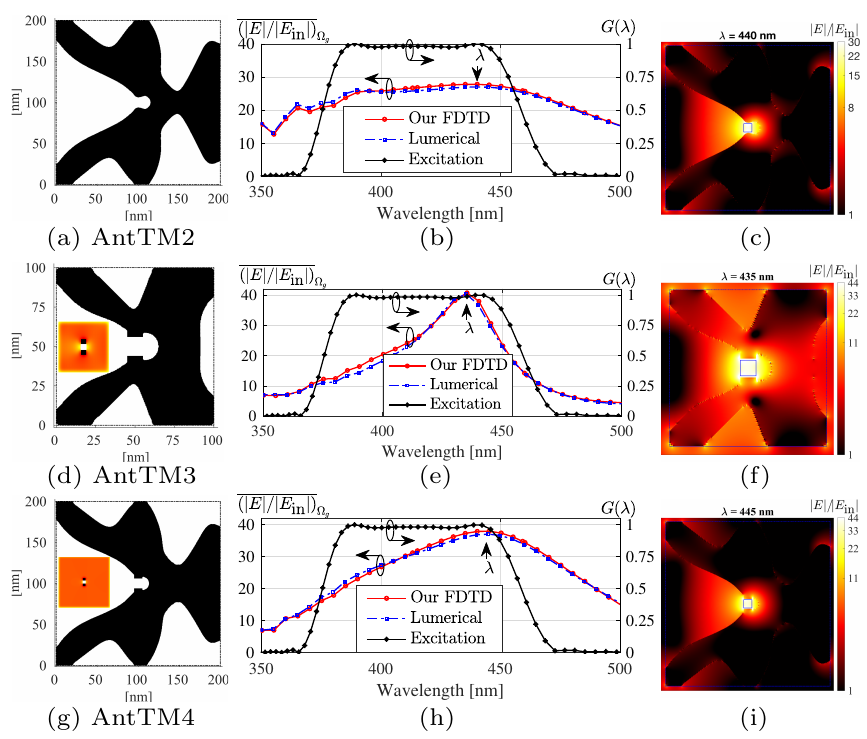}
\vspace{-10pt}
\caption{Topology (left), average field enhancement in $\Omega_g$ versus wavelength (centre), and field distribution at the wavelength of maximum average field enhancement for (a)-(c) AntTM2, (d)-(f) AntTM3, and (g)-(i) AntTM4, respectively. The second and third row show the effect of fixing the geometry around the gap $\Omega_g$ to relax the impact of a non-uniform sensitivity distribution. The colored insets in (d) and (g) show the design at iteration \#1, thus highlighting the geometry fixed around the gap $\Omega_g$.
}
\label{DesignAntTM2}
\end{figure}

The design obtained in Fig.\,\ref{Objective2DEx1_1IniResII} hits the boundary of the design domain, which suggests the need for a larger design space.
We double the size of the design domain to $200\!\times\!200$\,nm$^2$, and we solve the optimization problem, which now includes $319\,960$\,design variables.
The algorithm used $297$\,iterations to converge to the topology shown in Fig.\,\ref{DesignAntTM2}(a).
The new design, AntTM2, has more topological features compared to AntTM1.
We observe an additional vertical arm that evolved in the rear-side of the device, and the arm around $\Omega_g$ appears straight.
The performance of AntTM2 has improved at short wavelengths, and it attains a nearly flat response within the excitation window, as shown in Fig.\,\ref{DesignAntTM2}(b).
Fig.\,\ref{DesignAntTM2}(c) shows the field enhancement distribution of AntTM2 at the wavelength $440$\,nm.

\subsubsection{Effect of a fixed gap geometry}
Maximizing the energy in $\Omega_g$ suggests that the objective function is more sensitive to design variables close to $\Omega_g$ compared to those away from it.
We attempt to relax such a non-uniform distribution of the sensitivity and investigate its impact on the optimization results.
We fix the geometry region below and above $\Omega_g$ to silver with the same area as $\Omega_g$, and we solve the optimization problem.
Figs.\,\ref{DesignAntTM2}(d)-(f)  and Figs.\,\ref{DesignAntTM2}(g)-(i) show AntTM3 and AntTM4 which the algorithm produces when the design domain sizes $100\times100$\,nm$^2$ and $200\times200$\,nm$^2$ are used, respectively.
For AntTM3 and AntTM4, fixing the area around the gap boosts the average field enhancement to a maximum value of $40.6$ and $38$ at the wavelength $435$\,nm and $445$\, nm, respectively; AntTM1 and AntTM2 have maximum values of $27.8$ and $27.7$, respectively.
The new nanoantennas exhibit better performances at long wavelengths, however, the performance at short wavelengths stays essentially the same as in the previous cases.
These results indicate the challenges that plasmonic effects pose on optimizing nanoantennas near the surface plasmon frequency.

\subsubsection{Wideband optimization}
\begin{figure}
\includegraphics[trim = 0mm 0mm 0mm 0mm, clip ,width=1.0\columnwidth,draft=false]{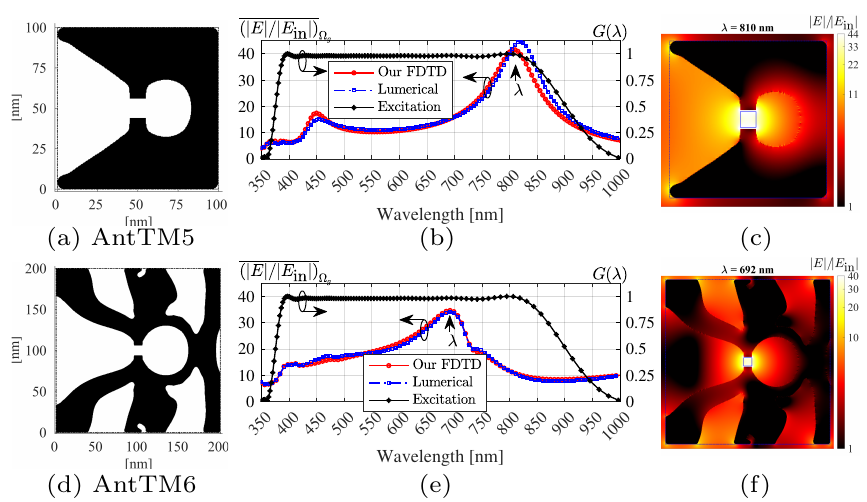}
\caption{Topology (left), average field enhancement in $\Omega_g$ versus wavelength (centre), and
field distribution at the wavelength marked in the second column of
(a)-(c) AntTM5, (d)-(f) AntTM6, respectively, which were optimized over the spectral window $375$--$900$\,nm. The domain size of AntTM5 and AntTM6 are $100\!\times\!100$\,nm$^2$ and $200\!\times\!200$\,nm$^2$, respectively.}
\label{DesignAntTMGapWBW}
\end{figure}

We combine the previous two investigations and pursue an attempt to optimize over a wider spectrum covering the wavelength window $375$--$900$\,nm.
That is, we use an excitation signal with a half-maximum bandwidth of $82$\% centered around $637.5$\,nm.
Figs.\,\ref{DesignAntTMGapWBW}(a)-(c) and Figs.\,\ref{DesignAntTMGapWBW}(d)-(f) show the topology, the average field enhancement, and the field distribution at the wavelength of the maximum performance of the new designs.
The new nanoantennas, called AntTM5 and AntTM6, show a wideband performance within the excitation spectrum.
An average field enhancement above $10$-fold is possible over the excitation window.
Moreover, the average field enhancement hits a maximum of $41.5$ and $34.5$ at the wavelength $810$\,nm and $690$\,nm for AntTM5 and AntTM6, respectively.
Here, we also observe the performance bias of the optimized nanoantennas towards long wavelengths.
Further investigations are needed to improve the broadband performance.

\subsection{2D, TE nanoantennas}
\begin{figure}
\includegraphics[trim = 0mm 0mm 0mm 0mm, clip ,width=1.0\columnwidth,draft=false]{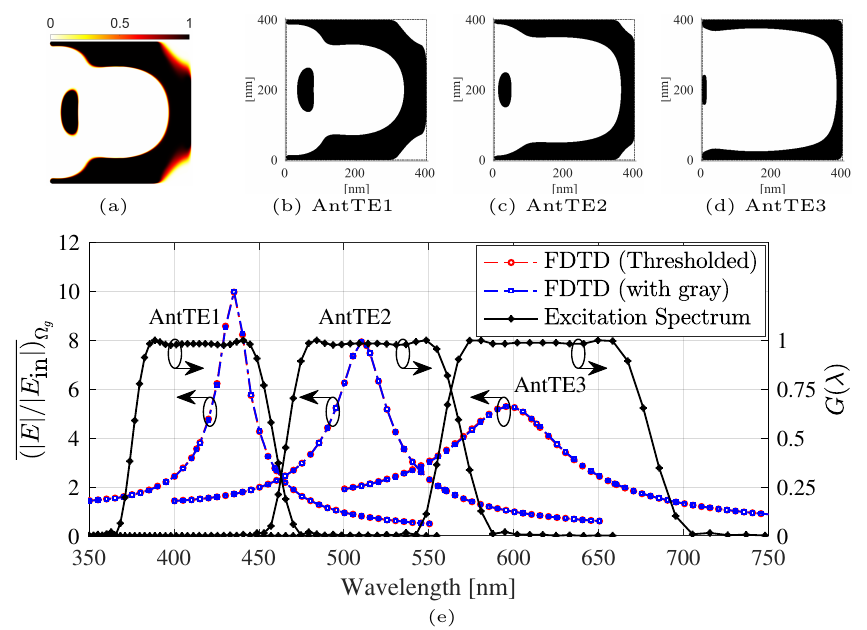}
\caption{TE topology optimized antennas over a design space $\Omega_d=400\times400$\,nm$^2$ with the observation domain $\Omega_g=20\times20$\,nm$^2$ centered in $\Omega_d$. (a) Final design obtained by the algorithm for AntTE1. Thresholded designs of (b) AntTE1, (c) AntTE2, and (d) AntTE3. (e) Average field enhancement inside $\Omega_g$ of the optimized antennas together with the excitation spectra.}
\label{Design2DTE}
\end{figure}

Surface plasmons develop at interfaces between metals and dielectrics only for TM polarization\cite{raether88surface,maier07plasmonics}.
Thus, plasmonic resonances in 2D structures are only possible for TM polarization.
This suggests that the artificial damping is not needed for optimization under TE polarization (the excitation is an $\mathcal{E}_x$ polarized plane wave), which is demonstrated in this section.
Here, the design variables are assigned to the out-of-plane permittivity components associated with the electric field $\mathcal{E}_x$, see Fig.\,\ref{Geom2D}(a).
We solve the optimization problem without using artificial damping.
Our numerical experiments show a monotonic increase of the objective function, and the algorithm exhibit no convergence problems. 
For brevity, we do not include these numerical evidences here, and we only show the optimization results.
Fig.\,\ref{Design2DTE} shows the optimization results of three nanoantennas optimized using three different excitation spectra.
We refer to these nanoantennas as AntTE1, AntTE2, and AntTE3.
For each design case, we shift the excitation spectrum by $100$\,nm. This allows us to investigate the effect of the electric size on the optimization.
For the three cases, we fix the size of the design domain $\Omega_d=400\times400$\,nm$^2$, the size of the observation domain $\Omega_g= 20\times20$\,nm$^2$, and we use $\Delta x \!=\! 2$\,nm.
A larger design domain was needed due to the absence of plasmonic resonances.
In Fig.\,\ref{Design2DTE}(a), we show the final design obtained by the algorithm for AntTE1.
We notice that for the TE cases, the final designs contain small amounts of intermediate materials. 
We evaluate the performance of the optimized antennas before and after thresholding these intermediate materials, around $\rho_\text{th} \!=\! 0.5$.
Fig.\,\ref{Design2DTE}(e) shows that the performance of the antennas is not sensitive to such remaining intermediate material, which explains why they are not removed by the algorithm.
For AntTE1, AntTE2, and AntTE3, the average field enhancement attains a maximum of $10.0$, $8.0$, and $5.3$, and occurs at the wavelength $435$\,nm, $510$\,nm, and $595$\,nm, respectively.
 The three structures have a paraboloidal shape reflector together with a standalone focusing segment, and the field enhancement is obtained as a result of a lensing effect, with $\Omega_g$ being the focus.
We remark that the absence of the plasmonic effects for the case of TE nanoantennas makes it hard to achieve field enhancement comparable to the TM nanoantennas with similar sizes.
Also, we note that the larger the electric size of the design domain, the better the results the algorithm can achieve for TE waves.

\subsection{3D antennas}
\begin{figure}
\includegraphics[trim = 0mm 1mm 0mm 0mm, clip ,width=1.0\columnwidth,draft=false]{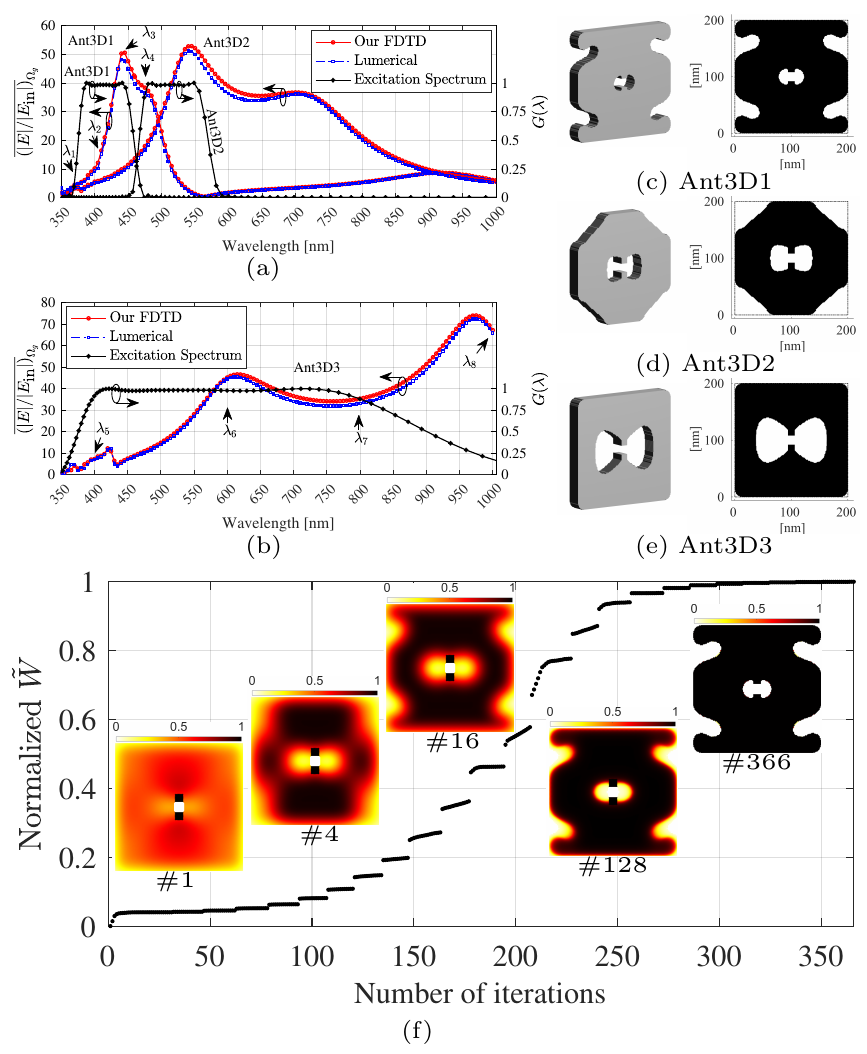}
\caption{Average field enhancements inside $\Omega_g$ together with the spectrum of excitation signals for (a) Ant3D1 and  Ant3D2,  (b) Ant3D3. Topologies of (c) Ant3D1, (d) Ant3D2, and (e) Ant3D3. (f) Progress of the objective function and some samples showing the development of Ant3D1.}
\label{EnergyAnt3D}
\end{figure}

As a final investigation, we use the developed method to optimize antennas in a 3D setup.
We extend the problem model given in Fig.\,\ref{Geom2D} to include a $30$\,nm thickness in the $x$-direction.
The design space has a volume $\Omega_d=200\!\times\!200\!\times\!30$\,nm${}^3$ with $\Omega_g \!=\!12\!\times\!12\!\times\!30$\,nm${}^3$, and we use $\Delta x\!=\! 2$\,nm.
Similar to the TM case, here we also fix the geometry region below and above $\Omega_g$ to silver with the same area as $\Omega_g$.
The discretized design domain includes $469\,341$\,design variables associated with its interior edges.
Using smaller design volumes would require finer discretization steps and longer simulation times, which increases the demand for memory resources.
We impose symmetry along the $x$-axis to enable antennas producible by current technologies.
That is, we optimize 3D\,antennas and aim for planar structures.
The excitation is an $\mathcal{E}_z$ polarized plane-wave propagating in the positive $x$\,axis.
We use the same setup and solve the optimization problem for three different wavelength excitation windows.
The first and second excitation spectra, shown in Fig.\,\ref{EnergyAnt3D}(a), have a half-maximum bandwidth of $20$\% centered around $413$\,nm and $513$\,nm, respectively.
The third excitation spectrum, shown in Fig.\,\ref{EnergyAnt3D}(b), has a half-maximum bandwidth covering the spectral window $375$--$900$\,nm. 
Each excitation spectrum results in a different topology which we name Ant3D1, Ant3D2, and Ant3D3, see Figs.\,\ref{EnergyAnt3D}(c)-(e).
Interestingly, we observe the increase of the figure-of-eight void area of the three nanoantennas, around the observation domain $\Omega_g$, as the excitation spectrum includes long wavelengths.
Fig.\,\ref{EnergyAnt3D}(f) shows the progress of the objective function and some snapshots of the intermediate designs for Ant3D1.

\begin{figure}
\vspace{-10pt}
\begin{center}
\includegraphics[trim = 0mm 0mm 0mm 0mm, clip ,width=1.0\columnwidth,draft=false]{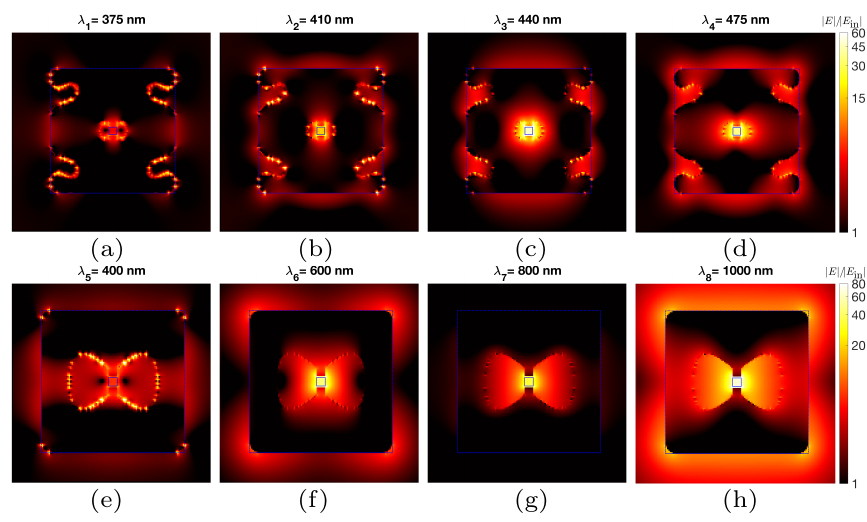}
\end{center}
\vspace{-15pt}
\caption{Field enhancements at the middle layer of (a)-(d) Ant3D1 and (e)-(h) Ant3D3 for some wavelengths marked in Fig.\,\ref{EnergyAnt3D}(a) and Fig.\,\ref{EnergyAnt3D}(b), respectively.}
\label{FieldAnt3D}
\end{figure}

Fig.\,\ref{EnergyAnt3D}(a) shows the average field enhancement of Ant3D1 and Ant3D2, and Fig.\,\ref{EnergyAnt3D}(b) shows the performance of Ant3D3.
Inside the observation domain $\Omega_g$, the nanoantennas Ant3D1, Ant3D2, and Ant3D3 exhibit a maximum average field enhancement of $50.5$, $52.8$, and $46.7$ at the wavelength $445$\,nm, $545$\,nm, and $615$\,nm, respectively.
Ant3D3 exhibits another peak of $74.0$ at $970$\,nm, which resides slightly outside the intended excitation spectrum, near-infrared wavelengths.
As in the 2D results, the optimized structures tend to exhibit a better field enhancement at long wavelengths.
Compared to the TE case, the presence of the plasmonic effects in the TM and 3D cases enabled much smaller nanostractures.
Figs.\,\ref{FieldAnt3D}(a)-(d) and Figs.\,\ref{FieldAnt3D}(e)-(h) show, respectively, the field distribution of Ant3D1 and Ant3D3 at some wavelengths, marked in Figs.\,\ref{EnergyAnt3D}(a)-(b).
The optimized structures are capable to maximize the electric energy at the observation domain $\Omega_g$.
At short wavelengths, however, we observe high-field localizations near their boundaries, which indicates strong plasmonic effects that are responsible for the decrease in the achieved performances.
Similar to the 2D TM case, further investigations are needed to obtain a balanced performance over the wavelength window of interest. 

\section{Conclusion}
\label{S:Con}
We introduced a density-based topology optimization approach to design plasmonic dispersive nanoantennas.
Our approach is based on Maxwell's equations in the time-domain, and we use the Drude model, which can fit the material dispersion of metals and conductive polymers, as well as epsilon-near-zero materials, such as conductive oxides.
For the TM and the 3D setups, the interpolation between metallic and dielectric phases results in high field-localization associated with plasmonic effects, which prevent the algorithm from converging to well-performing designs.
Guided by dispersion diagrams of metal-dielectric interfaces, we proposed an artificial damping approach to suppress the field-localization during the optimization process, which enables the algorithm to converge to good designs.
For the TE setup, artificial damping is not needed and the algorithm encounters no convergence issues.
Various setups for narrowband and wideband optimization are presented, resulting in novel 2D and 3D nanoantenna designs with outstanding performances.
Our method opens new opportunities for the automatic design and optimization of dispersive nanophotonic structures with broadband optical response for nanoplasmonics, nonlinear optics, epsilon-near-zero materials, ultrafast photonics, or integrated optics.

\section*{Acknowledgement}
The computations were performed on resources provided by the Swedish National Infrastructure for Computing (SNIC) at HPC2N center; the central computing cluster operated by Leibniz University IT Services (LUIS); and the North-German Supercomputing Alliance (HLRN). A.C.L. acknowledges the Bundesministerium für Buldung und Furschung (German Federal Ministry of Education and Research) under the Tenure-Track Program, and the Deutsche Forschungsgemeinschaft (DFG, German Research Foundation) under Germany’s Excellence Strategy within the Cluster of Excellence PhoenixD (EXC 2122, Project ID 390833453).

\bibliography{IEEEabrv,MyAbrv,references}

\end{document}